\documentclass[%
superscriptaddress,
preprint,
amsmath,amssymb,
aps,
]{revtex4-1}
\usepackage{graphicx}
\usepackage{dcolumn}
\usepackage{bm}
\usepackage{amsmath}
\usepackage{amsthm}
\usepackage{booktabs}
\usepackage{float}
\usepackage{braket}
\usepackage[utf8]{inputenc}
\begin{document}


\title{Coupling of phase transition, anharmonicity, and thermal transport in CaSnF$_6$}

\author{Daxue Hao}
\affiliation{College of Physics Science and Technology, Yangzhou University, Jiangsu 225009, China}

\author{Hao Huang}
\affiliation{Advanced Copper Industry College, Jiangxi University of Science and Technology, Yingtan 335000,China}

\author{Geng Li}
\affiliation{China Rare Earth Group Research Institute, Shenzhen, Guangdong 518000, China}
\affiliation{Key Laboratory of Rare Earths, Ganjiang Innovation Academy, Chinese Academy of Sciences, Ganzhou, 341000, China}

\author{Yu Wu}
\email{wuyu@njnu.edu.cn}
\affiliation{Micro- and Nano-scale Thermal Measurement and Thermal Management Laboratory, School of Energy and Mechanical Engineering, Nanjing Normal University,Jiangsu,
Nanjing 210023, China}

\author{Shuming Zeng}
\email{zengsm@yzu.edu.cn}
\affiliation{College of Physics Science and Technology, Yangzhou University, Jiangsu 225009, China}


\date{\today}

\begin{abstract}
Understanding the coupling between structural phase transitions and thermal transport is essential for designing functional materials with tunable properties. Here, we investigate this interplay in CaSnF$_6$ by combining first-principles calculations with a machine-learned neuroevolution potential that enables large-scale molecular dynamics simulations across a wide temperature range. The simulations accurately capture the first-order structural phase transition and associated lattice dynamics. We show that the negative thermal expansion originates from low-energy rigid unit modes involving cooperative rotations of corner-sharing [CaF$_6$]$^{4-}$ and [SnF$_6$]$^{2-}$ octahedras, which induce bond-angle bending and volume contraction. At the same time, strong anharmonicity, dominated by four-phonon scattering, plays a central role in suppressing lattice thermal conductivity ($\kappa_L$). Crucially, non-equilibrium simulations reveal a pronounced non-monotonic anomaly in $\kappa_L$ near the phase transition, deviating from the conventional $\sim 1/T^{\alpha}$ behavior and providing direct transport evidence of lattice reconstruction. These results establish a unified mechanism linking lattice geometry, anharmonic vibrational dynamics, and thermal transport, and highlight the potential of machine-learned potentials for bridging atomic-scale phase transitions with macroscopic transport properties.

\end{abstract}



\maketitle

\section{Introduction}
Thermal transport properties are key performance indicators for the design and application of functional materials, playing important roles in various fields such as electronic thermal management and thermoelectric energy conversion\cite{qian2021phonon,yan2022high}. Lattice thermal conductivity ($\kappa_L$) can be calculated by first-principles calculations of phonon dispersion relations and phonon-phonon scattering rates combined with solving the Boltzmann transport equation (BTE), or through molecular dynamics (MD) simulations of heat current autocorrelation functions using the fluctuation-dissipation theorem (FDT)\cite{carbogno2017ab,kang2017first}. The former approach can decompose the contributions of different phonon modes (frequency, wave vector) to $\kappa_L$, revealing the dominant phonon frequency ranges and scattering mechanisms governing heat transport. While the latter includes full-order anharmonicity, it suffers from significant size effects requiring large systems to reduce errors, and both methods are limited by their finite accuracy or high computational costs. In recent years, various machine learning potentials (MLPs) have provided new approaches for calculating thermal properties of materials. Many MLPs such as Behler-Parrinello neural-network potential (BPNNP), Gaussian approximation potential (GAP), spectral neighbor analysis potential (SNAP), moment tensor potential (MTP), deep potential (DP), and atomic cluster expansion (ACE) have been widely applied in MD simulations\cite{behler2007generalized,bartok2010gaussian,caro2019optimizing,byggmastar2022simple,thompson2015spectral,novikov2020mlip,wang2018deepmd,drautz2019atomic}. The neuroevolution potential (NEP) developed by Fan et al. demonstrates outstanding accuracy and leading computational efficiency compared to other programs, showing unique advantages\cite{fan2021neuroevolution,fan2022improving,fan2022gpumd}.

Negative thermal expansion (NTE) materials have shown significant application potential in thermal management and heat transport fields due to their unique volume contraction characteristics and anomalous thermal properties\cite{liang2021negative}. NTE behavior typically originates from special lattice dynamics mechanisms within materials, such as transverse vibrational modes, atomic flipping motions, or anharmonic interactions. These materials can compensate for dimensional changes in positive expansion materials, thereby regulating the overall thermal expansion coefficient in composites and greatly improving dimensional stability of devices, which is particularly crucial for microelectronics, optical devices, and aerospace equipment\cite{oddone2017composites}. In terms of heat transport, NTE materials exhibit significantly different $\kappa_L$ characteristics compared to conventional materials. Their low $\kappa_L$ and glass-like behavior provide new possibilities for thermal insulation and energy management applications. For instance, typical NTE materials like ScF$_3$ display anomalous $\kappa_L$ behavior due to their negative Gr\"{u}neisen parameters\cite{tang2023anomalous,tang2024effects}, microscopically reflecting the suppression of heat conduction by lattice vibrations. Studies show that four-phonon (4ph) interactions and thermal expansion effects play important roles in these materials, significantly affecting their $\kappa_L$. Additionally, the thermal insulation behavior of other NTE materials like CuSCN\cite{shen2020ultralow} suggests great potential in phonon heat transport regulation. These studies not only reveal the complex relationship between thermal expansion and $\kappa_L$ but also provide theoretical support for developing new thermal management materials.

Recently, Ezra Day-Roberts et al. reported the double perovskite fluoride CaSnF$_6$ as a ferroaxial compound that undergoes a phase transition from a cubic phase to a phase with a axial moment~\cite{day2025piezoresistivity}, while Qilong Gao et al. proposed a low-cost synthesis method \cite{gao2023new}. However, studies on its thermal properties remain scarce. Given the challenges of traditional first-principles calculations in investigating thermal expansion and phase transition properties, this study combines first-principles calculations with the NEP, employing the BTE and homogeneous nonequilibrium molecular dynamics (HNEMD) \cite{j2007statistical,fan2019homogeneous} methods to systematically investigate the evolution of thermal transport properties and phase transition behavior of CaSnF$_6$ in the temperature range of 75-600 K. Through the GPUMD\&NEP package, a high-accuracy potential function was constructed via multiple rounds of active learning, with its training set covering both high-temperature and low-temperature phase structures of CaSnF$_6$. The reliability of the potential was confirmed by verifying energy-force consistency with DFT calculations. Using this potential to perform MD simulations across the aforementioned temperature range and extracting force constants, the calculated $\kappa_L$ at 300K considering three-phonon (3ph) and 4ph contributions from BTE solutions were 7.02 W/mK and 3.49 W/mK respectively, highlighting the crucial role of quartic anharmonicity. Calculations considering thermal expansion effects using unit cells constructed from NEP-MD trajectories yielded results of 5.23 W/mK and 2.46 W/mK. HNEMD calculations showed thermal conductivity of 3.78 W/mK without considering thermal expansion, decreasing to 2.93 W/mK when thermal expansion effects were included, confirming the enhanced phonon scattering due to volume contraction. Notably, the constructed NEP potential successfully predicted the negative thermal expansion coefficient ($\alpha_v = -14.67\times10^{-6}$ K$^{-1}$) of the high-temperature phase of CaSnF$_6$, showing excellent agreement with experimental values and validating its reliability in complex lattice dynamics simulations.

\section{methodology}
In this study, density functional theory (DFT) \cite{hohenberg1964density,kohn1965self} calculations were performed using the VASP \cite{kresse1993ab,kresse1994ab,kresse1996efficiency,kresse1996efficient} package with projector augmented wave (PAW) pseudopotentials \cite{kresse1999ultrasoft}. The electron exchange-correlation functional was described by PBEsol \cite{perdew1996generalized}, and a plane-wave basis set cutoff energy of 600~eV was used. The Brillouin zone was sampled using a $\Gamma$-centered $7\times7\times7$ $k$-mesh. Structural optimizations were converged to a total energy of $10^{-7}$~eV and atomic forces of $0.01$~eV/\AA. To investigate thermal transport properties at finite temperatures, $ab$ $initio$ molecular dynamics (AIMD) simulations were carried out in an NVT ensemble on a $3\times3\times2$ supercell (144 atoms) with a time step of 1~fs and a total simulation time of 20~ps, covering a temperature range of 250--600~K (Fig.~S1). Interatomic force constants (IFCs) were extracted from the AIMD trajectories using the temperature-dependent effective potential (TDEP) method \cite{knoop2024tdep}. The cutoff radii for the second-, third-, and fourth-order force constants were set to 8~\AA, 6~\AA, and 4~\AA, respectively \cite{hellman2013temperature,hellman2013temperature14,romero2015thermal}. The ShengBTE \cite{li2012thermal,li2014shengbte,han2022fourphonon} package was used to self-consistently solve the linearized phonon Boltzmann transport equation (BTE) through an iterative method, avoiding the underestimation errors potentially introduced by the relaxation time approximation (RTA).

The expression for the thermal conductivity tensor can be given in the energy transport form as:
\begin{equation}
\kappa_L^{\mu\nu} = \frac{1}{N V} \sum_{\lambda} C_\lambda v_\lambda^\mu F_\lambda^\nu,
\label{eq:bte}
\end{equation}
where $C_\lambda$ is the modal specific heat corresponding to mode $\lambda = (\mathbf{q}, s)$, $v_\lambda^\mu$ is the component of the phonon group velocity in the $\mu$ direction, and $F_\lambda^\nu$ represents the deviation of the distribution function for this mode in the $\nu$ direction under perturbation, which is solved by the following iterative form:
\begin{equation}
F_\lambda^{(n+1)} = \tau_\lambda^{(n)} \left[ v_\lambda^\nu + \sum_{\lambda'} \Lambda_{\lambda\lambda'} F_{\lambda'}^{(n)} \right],
\end{equation}
where $\tau_\lambda$ is the effective relaxation time, $\Lambda_{\lambda\lambda'}$ represents the linear scattering rate coefficient matrix for inter-mode coupling, and $n$ denotes the iteration step. ShengBTE starts from an initial RTA solution and progressively updates the deviation distribution through the above equations until a self-consistent convergence criterion is met, thereby yielding the anharmonic thermal conductivity tensor across the entire wave vector range. The $q$-mesh density is a critical parameter in thermal conductivity calculations. We performed detailed convergence tests of the thermal conductivity with respect to different $q$-mesh densities (Fig.~S2), ultimately determining that a $16\times16\times16$ $q$-mesh should be used.

The training set for the neuroevolution potential (NEP) was constructed based on the primitive cells of both the low-temperature (rhombohedral) and high-temperature (cubic) phases of CaSnF$_6$. Initial structures were generated by applying $\pm3\%$ lattice distortions and $\pm0.03$~\AA\ random atomic displacements to an optimized $3\times3\times3$ supercell (216 atoms). High-precision single-point energy calculations used an automatic $k$-spacing of 0.03~\AA$^{-1}$ (i.e., $0.03\times2\pi/\text{\AA}$). An active learning strategy was adopted to optimize the potential function: first, NEP-driven MD simulations were performed across multiple temperature regions. Subsequently, 100 representative configurations were selected using the farthest point sampling (FPS) method for DFT single-point calculations, and configurations with larger prediction errors were iteratively added to the training set. The final training set excluded configurations where atomic forces exceeded $\pm15$~eV/\AA, ensuring that the data concentrated on equilibrium states relevant to thermal transport.


Molecular dynamics (MD) calculations of $\kappa_L$ have long relied on the Green-Kubo (GK) linear response theory, whose core formula (Eq.~\ref{eq:gk}) relates thermal conductivity to the time integral of the equilibrium heat flux autocorrelation function:
\begin{equation}
\kappa_L^{\mu\nu}(t) = \frac{V}{k_B T^2} \int_0^t dt' \left\langle Q_\mu(0) Q_\nu(t') \right\rangle.
\label{eq:gk}
\end{equation}
Here, \(\kappa_L^{\mu\nu}(t)\) represents the time-dependent thermal conductivity tensor, with indices \(\mu\) and \(\nu\) corresponding to spatial directions. The integral of the autocorrelation function \(\langle Q_\mu(0) Q_\nu(t') \rangle\) over time \(t'\) captures the temporal decay of equilibrium heat flux fluctuations. The parameters \(V\), \(k_B\), and \(T\) denote the system volume, Boltzmann constant, and temperature, respectively.

To address the computational efficiency limitations of equilibrium methods (like GK), the homogeneous non-equilibrium molecular dynamics (HNEMD) method introduces a driving force \(F_e\) through a modified momentum equation:
\begin{equation}
\frac{d p_i}{dt} = F_i + C(q,p)F_e,
\end{equation}
where \(C(q,p)\) is a phase space function that depends on the coordinates \(q\) and momenta \(p\) of all particles. This driving term \(C(q,p)F_e\) simulates a thermodynamic force that perturbs the system into a non-equilibrium steady state. The function \(C(q,p)\) is constructed to ensure phase space incompressibility, as dictated by the Liouville equation.
Under steady-state conditions, the non-equilibrium heat flux response is linearly proportional to the driving force \(F_e\):
\begin{equation}
\frac{\langle J^\mu(t) \rangle_{\text{ne}}}{T V} = \sum_{\nu} \kappa_L^{\mu\nu} F_e^\nu.
\label{eq:hnemd}
\end{equation}
In this expression, \(\langle J^\mu(t) \rangle_{\text{ne}} \) denotes the time-averaged heat flux in the \(\mu\) direction under the influence of \(F_e\), while \(\kappa_L^{\mu\nu}\) quantifies the thermal conductivity tensor components. The summation over \(\nu\) reflects the tensor nature of heat conduction in anisotropic systems.

To systematically investigate the effect of thermal expansion on thermal transport properties, this study comprehensively utilized both the BTE and HNEMD methods for complementary validation and trend analysis.
For achieving better statistical accuracy at the spatial scale, the model used for HNEMD calculations was a $10\times10\times10$ supercell (totaling 8000 atoms) constructed from the optimized primitive cell of the high-temperature phase. The reliability of this model within the studied temperature range was verified through size convergence tests (see Supplementary Fig.~S3). A simulation time step of 1~fs was set. For calculations not considering thermal expansion effects, the system was first equilibrated for 1~ns under fixed volume (NVT ensemble), followed by sampling of 10~ns of heat flux data for steady-state thermal conductivity calculation. In the case considering thermal expansion, the simulation procedure first involved equilibrating the system for 1~ns under isothermal-isobaric (NPT) conditions to obtain the equilibrium volume. Subsequently, the structure was frozen, and the simulation transitioned into an NVT ensemble for sampling heat flux trajectories, with a sampling time also of 10~ns. To enhance statistical precision, each temperature point was independently simulated 5 times, and the final results were averaged across these replicas, with the error fluctuations demonstrated in Supplementary Figs.~S4 and S5.

For NEP-combined BTE calculations, the corresponding equilibrium volume at each temperature was first obtained through NPT ensemble NEPMD simulations. Subsequently, the primitive cell structure was reconstructed, and NVT-MD trajectories were generated. These trajectories were then used to construct TDEP force constants, which were in turn coupled with ShengBTE to solve the lattice thermal conductivity. This methodological framework is consistent with the previously described AIMD-TDEP-ShengBTE workflow but, by utilizing the efficient NEP, significantly reduces computational costs, enabling the precise capture of anharmonic effects over large scales and long timescales. Furthermore, to analyze the structural phase transition behavior of CaSnF$_6$, starting from the low-temperature rhombohedral structure, NPT-MD simulations were performed on its 8000-atom supercell within the 100--300~K temperature range. By monitoring the temperature evolution of lattice parameters and system potential energy, a clear structural discontinuity was observed near 143~K, confirming the phase transition characteristics between the rhombohedral and cubic phases as reported experimentally.

\section{Results and Discussions}
\begin{figure}[ht]
	\centering
	\includegraphics[width=16cm]{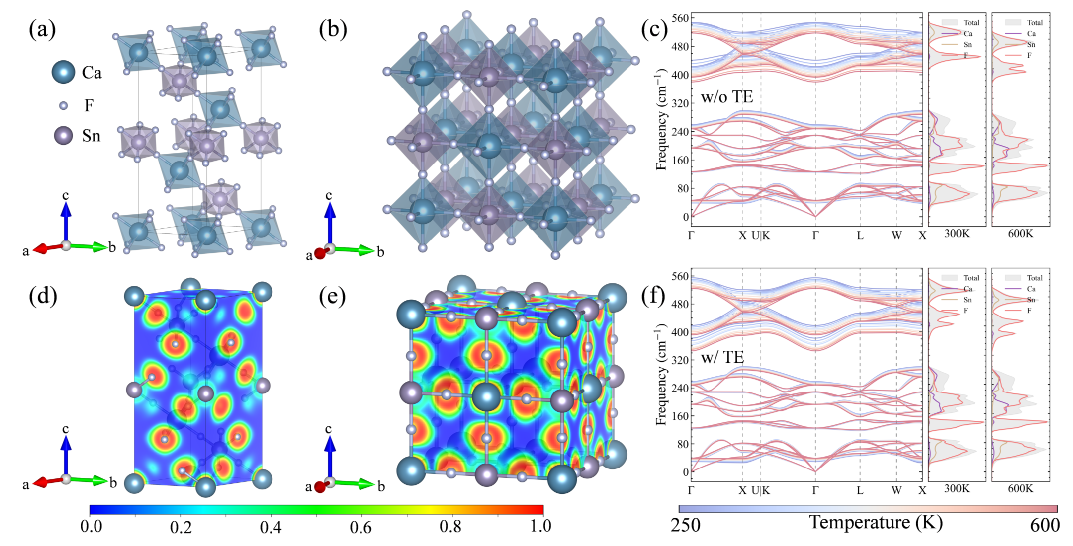}
	\caption{The crystal structures of (a) low temperature phase (rhombohedral) and (b) high temperature phase (cubic) of CaSnF$_6$. (d) and (e) are the electron localization function (ELF) plot of rhombohedral and cubic CaSnF$_6$. The phonon dispersion of cubic CaSnF$_6$ (c) without thermal expansion (w/o TE) and (f) with thermal expansion (w/ TE). The color scale (blue to red) represents the temperature variation from 250 K to 600 K.
        The phonon dispersions are calculated from NEP-MD trajectories.}
\label{fig:structure}
\end{figure}

As shown in Figs.~\ref{fig:structure}(a)-(b), CaSnF$_6$ exhibits significant structural differences between its high-temperature and low-temperature phases. The high-temperature phase crystallizes in the cubic system (space group $Fm\overline{3}m$, No. 225) with a conventional (primitive) cell 32 (8) atoms, forming a face-centered cubic framework composed of corner-sharing [SnF$_6$]$^{2-}$ and [CaF$_6$]$^{4-}$ octahedra. Below 143 K, the material transforms into a rhombohedral phase (space group $R\overline{3}$, No. 148). Bader charge analysis (Table I) quantitatively characterizes the bonding nature: The substantial deviation of Sn's actual charge (+2.51$|e|$) from its formal charge (+4$|e|$) confirms polar covalent character in Sn-F bonds, while the intermediate Ca charge (+1.63$|e|$ vs formal +2$|e|$) indicates ionic bonding nature for Ca-F bonds. Crucially, fluorine atoms carry an effective charge of -0.70$|e|$, significantly deviating from the nominal -1$|e|$, indicating partial electron delocalization and a polar covalent bonding character. This electronic feature provides a basis for modified interatomic force constants and enhanced lattice anharmonicity. Within the temperature range of 250-600 K, the low-frequency region, which dominated by transverse vibrations of F atoms, exhibits an unusual combination of phonon frequency upshift and linewidth broadening [see Figs.1(c)-(f)]. Notably, the low-frequency optical branches show clear hardening alongside a significant increase in linewidth, suggesting that mode-specific force constant enhancement coexists with strong anharmonic phonon scattering. In contrast, the mid-to-high frequency range shows conventional softening. This paradoxical competition reveals that in NTE systems, lattice contraction can selectively stiffen certain vibrational modes, while thermal effects simultaneously enhance anharmonic phonon-phonon interactions. These features elucidate the microscopic origin of the apparent ``structural contraction-dynamical softening" interplay in NTE materials\cite{PhysRevLett.107.195504}.

A high-quality interatomic potential is crucial for subsequent calculations of phase transitions and transport properties. Figures S6-7 compares the physical quantities predicted by the trained NEP model with those calculated using DFT methods. The RMSEs for energy, force, virial and stress on the training set are 0.682 meV/atom, 33.069 meV/\AA, 3.746 meV/atom, and 32.337 MPa, respectively, with corresponding test set values of 1.253 meV/atom, 44.260 meV/\AA, 4.196 meV/atom, and 36.669 MPa. The small RMSE values for both sets demonstrate the accuracy and reliability of our trained NEP model. To further validate the NEP model's reliability, we compared the finite-temperature phonon dispersion relations of cubic CaSnF$_6$ (Fig. S8), noting slight deviations around 400 cm$^{-1}$ in the optical branches, which is consistent with common limitations in many-body interaction simulations. However, thermal transport is dominated by low-frequency phonons, as their longer relaxation times and group velocities enable greater heat carriage. Specifically, for CaSnF$_6$, cumulative $\kappa_L$ analysis reveals that over 80\% of the $\kappa_L$ is contributed by phonons below 100 cm$^{-1}$. Our trained NEP model reproduces excellent agreement with DFT methods in the acoustic branches, thus we can expect reliable predictions of thermal transport properties for CaSnF$_6$ from NEP model.

The phonon dispersion [Figs.~\ref{fig:structure}(c)-(f)] with (w/TE) and without (w/o TE) considering thermal expansion both show no imaginary frequencies, confirming the dynamical stability of the corresponding structures. Additionally, Table S1 presents the independent elastic constants, which satisfy the Born stability criteria, indicating that CaSnF$_6$ is also mechanically stable. While the overall trends of the two spectra are consistent, they exhibit subtle differences opposite to conventional materials: despite negative thermal expansion, w/TE phonon frequencies are generally slightly lower than w/o TE due to enhanced anharmonicity of atomic vibrations rather than lattice expansion. This frequency reduction is more pronounced in the high-frequency region ($>360$ cm$^{-1}$) of optical branches, particularly for F$^-$ ion localized vibration modes that are most sensitive to temperature changes, accompanied by slight shifts near high-symmetry points. Phonon density of states reveals a collective downward shift of high-frequency spectra with increasing temperature, where F atoms dominate contributions across all frequencies, Sn atoms primarily govern low-frequency vibrations ($<100$ cm$^{-1}$), and Ca atoms mainly influence the mid-frequency range ($150-300$ cm$^{-1}$). Notably, the high-frequency F vibration modes split from two peaks into three when considering temperature effects, reflecting symmetry breaking in the local chemical environment induced by lattice contraction upon heating. This anomalous ``contraction-induced phonon splitting" stems from subtle reconstruction of F coordination fields. These changes demonstrate that temperature effects significantly alter lattice dynamics through modified interatomic interactions, particularly for high-frequency optical modes dependent on short-range interactions. Although lattice contraction would normally be expected to cause phonon hardening, the observed frequency reduction suggests dominant anharmonic effects that may further modify phonon lifetimes and scattering mechanisms, thereby profoundly influencing thermal transport properties at elevated temperatures.

\begin{figure}[ht]
    \centering
    \includegraphics[width=16cm]{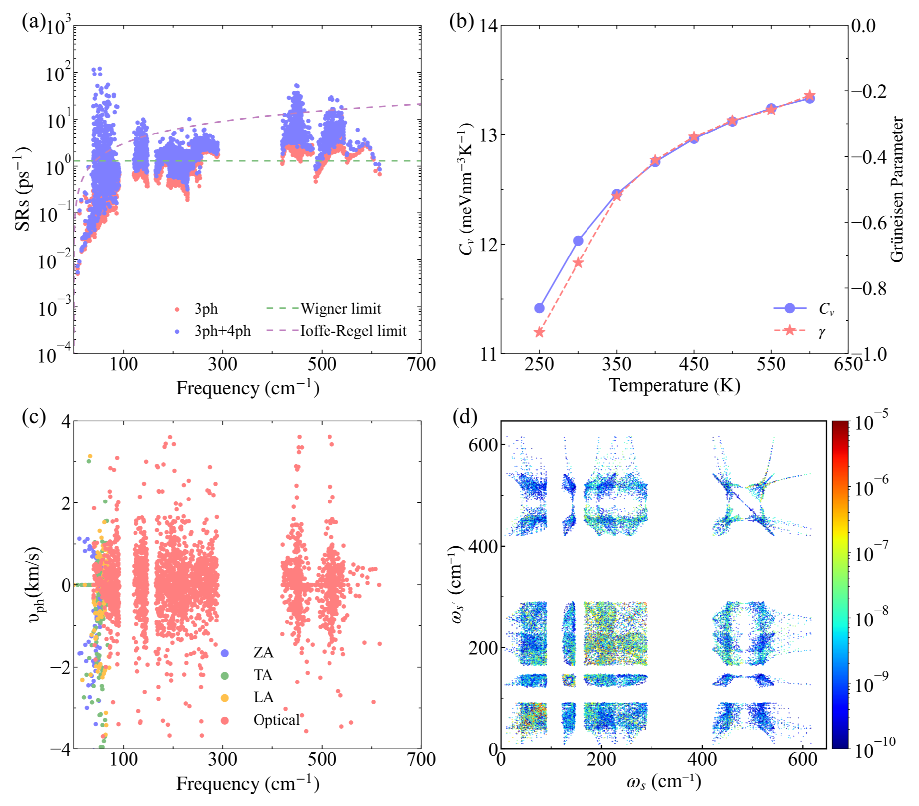}
    \caption{
        Phonon transport characteristics of CaSnF$_6$ obtained from AIMD trajectories.
        (a) Mode-resolved phonon scattering rates at 300 K.
        (b) Temperature-dependent heat capacity (left axis) and mode Gr\"uneisen parameters (right axis) over 250--600 K.
        (c) Phonon group velocities at 300 K.
        (d) Frequency-resolved contribution to the cross-plane thermal conductivity $\kappa_c$ at 300 K, projected onto phonon mode pairs ($\omega_s$, $\omega_{s'}$).
    }
    \label{fig:phonon_properties}
\end{figure}

Phonon scattering rates in CaSnF$_6$, including 3ph processes and 3ph+4ph contributions, are shown in Fig.~\ref{fig:phonon_properties}. In the low-frequency regime ($<50$ cm$^{-1}$), phonon scattering rates remain low and are dominated by 3ph processes, indicating relatively weak anharmonic interactions. With increasing frequency, scattering rates rise rapidly and approach the Ioffe-Regel limit\cite{beltukov2013ioffe}, signaling a breakdown of well-defined phonon quasiparticles due to shortened mean free paths and enhanced anharmonicity. In the mid- to high-frequency range, 4ph processes become dominant over multiple frequency intervals, with pronounced spectral peaks. These higher-order scattering channels substantially suppress the $\kappa_L$ compared to 3ph-only predictions (the difference between the two is more than double, as shown in Fig.S10). Notably, some 3ph+4ph scattering rates exceed the Ioffe-Regel limit, suggesting a crossover from coherent phonon transport to incoherent, diffusion-dominated vibrational dynamics. Therefore, calculating the $\kappa_L$ based on the phonon picture may introduce some errors.

The heat capacity ($C_v$) increases nonlinearly over 250-600 K, approaching saturation at high temperature in agreement with the Debye model. Concurrently, the negative Gr\"uneisen parameter weakens from $-0.94$ at 250 K to $-0.21$ at 600 K, indicating that anharmonic vibrations progressively diminish volumetric effects at elevated temperatures. Combined with group velocity ($v_{ph}$) analysis in the Fig.2(c) and Fig.~S9, a slight reduction in low-frequency group velocities with increasing temperature is observed. This trend directly correlates with the dual characteristics of ``frequency-upshift and linewidth-broadening" identified in the phonon spectra: lattice contraction induces phonon hardening (blue-shift) via the negative Gr\"uneisen effect, while temperature-induced anharmonicity concurrently reduces the slope of dispersion curves, resulting in decreased effective $v_{ph}$. As shown in Fig. S9, with increasing temperature, the phonon scattering rates and $v_{ph}$ in the low-frequency region shift toward higher frequencies, whereas those in the high-frequency region shift toward lower frequencies. This behavior is consistent with the coexistence of phonon hardening and softening observed in the phonon spectra. Notably, in the presence of thermal expansion (w/TE), the phonon scattering rates are significantly enhanced in the low-frequency regime, which plays a dominant role in suppressing the $\kappa_L$.

\begin{figure}[ht]
    \centering
    \includegraphics[width=16cm]{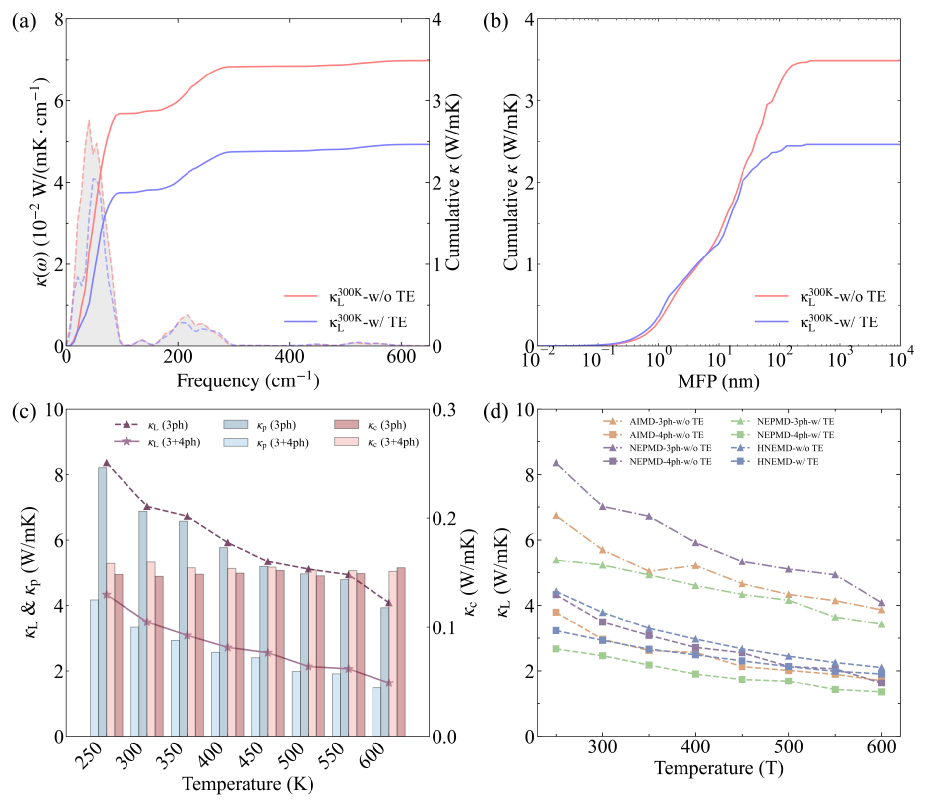}
    \caption{
  Thermal transport properties of CaSnF$_6$.
(a) Spectral and cumulative thermal $\kappa_L$ as a function of phonon frequency.
(b) Cumulative $\kappa_L$ as a function of phonon mean free path (MFP).
(c) Decomposition of $\kappa_L$ into particle-like ($\kappa_p$) and glass-like ($\kappa_c$) components, with $\kappa_L = \kappa_p + \kappa_c$.
(d) Comparison of $\kappa_L$ obtained from different theoretical approaches.
Calculations in (a-c) are based on NEPMD trajectories.
    }
    \label{fig:kappa}
\end{figure}

\begin{figure}[ht]
    \centering
    \includegraphics[width=12cm]{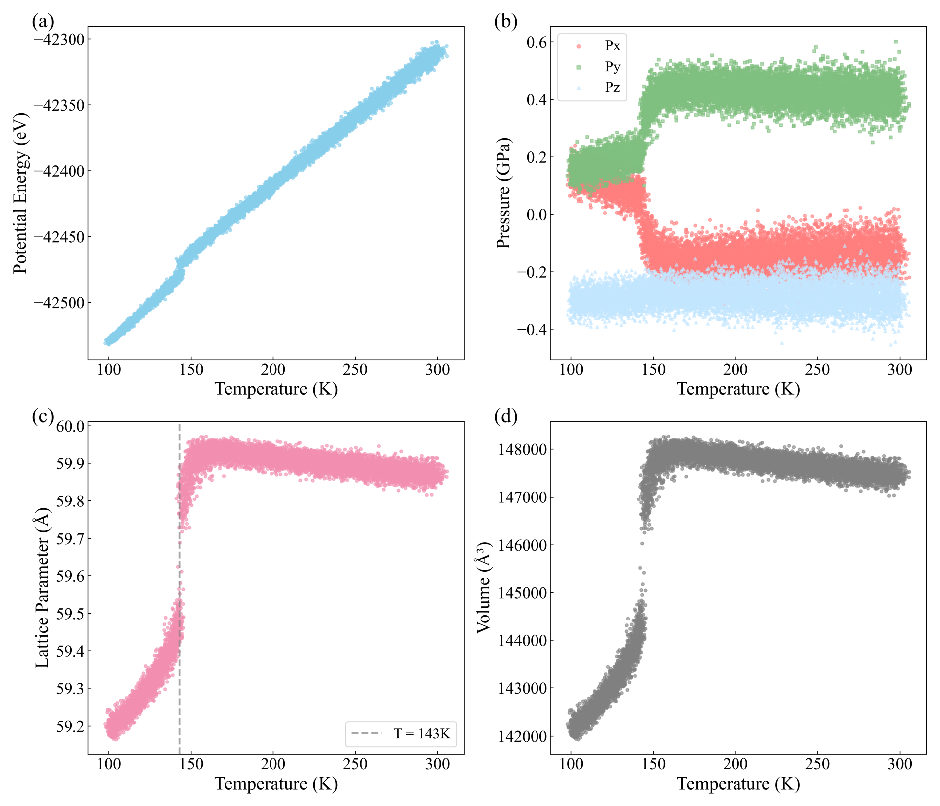}
    \caption{
(a) System potential energy as a function of temperature, showing an abrupt change near 143 K indicative of a first-order structural phase transition.
(b) Temperature-dependent pressure along the three lattice directions.
(c) Lattice parameter along the a-axis as a function of temperature, exhibiting positive thermal expansion below 143 K and negative thermal expansion above 143 K.
(d) Total supercell volume as a function of temperature.
    }
    \label{fig:phase_transition}
\end{figure}

Frequency-dependent analysis reveals that low-frequency phonons below ($100$ cm$^{-1}$) contribute 80\% of the $\kappa_L$ (Fig.~\ref{fig:kappa}a), confirming the dominant role of this frequency range identified in prior phonon spectrum analysis. The cumulative mean free path (MFP) curve (Fig.~\ref{fig:kappa}b) shows maximum cumulative $\kappa_L$ at 335 nm, consistent with the long MFP characteristic of low-frequency phonons. Temperature-dependent studies (Fig.~\ref{fig:kappa}c) demonstrate continuous decrease in $\kappa_L$ and $\kappa_p$ with rising temperature, aligning with enhanced phonon scattering mechanisms, while the glass-like thermal conductivity $\kappa_c$ remains relatively stable. This phenomenon correlates with the previously observed evolution of Gr\"uneisen parameters and slight reduction in $v_{ph}$: anharmonic vibrations diminish volumetric effects while sustaining incoherent heat transport. Frequency-resolved $\kappa_c$ further demonstrates that primary contributions originate from the diagonal region where $\omega_s \approx \omega_{s'}$ [shown in Fig.2(d)]. As shown in Fig.~3(c) and Fig.~S10(a-c), inclusion of 4ph scattering results in a pronounced reduction in $\kappa_L$, lowering the 3ph+4ph values to less than half of the 3ph-only results, regardless of thermal expansion. This trend is fully consistent with the corresponding scattering rate analysis. Figure~3(d) compares $\kappa_L$ values obtained from phonon-based (AIMD, NEPMD+BTE) and heat-flux-based (HNEMD) approaches, with and without thermal expansion. Without thermal expansion, NEPMD and HNEMD overestimate the $\kappa_L$ relative to fully first-principles results. Inclusion of 4ph scattering brings NEPMD into close agreement with AIMD. In contrast, accounting for thermal expansion leads to a pronounced reduction in $\kappa_L$, driven by enhanced low-frequency scattering. In CaSnF$_6$, negative thermal expansion leads to lattice contraction that would, in principle, enhance $v_{ph}$ via phonon hardening. This effect, however, is overwhelmed by increased anharmonicity and temperature-driven scattering. As a result, the $\kappa_L$ decreases from 3.49 to 2.46 W m$^{-1}$ K$^{-1}$ upon inclusion of thermal expansion (Fig.~\ref{fig:kappa}(a)) at 300 K, demonstrating that reduced phonon lifetimes dominate over the gain in group velocity.

To probe the structural phase transition in CaSnF$_6$, we performed NPT molecular dynamics simulations on an 8,000-atom supercell derived from the low-temperature rhombohedral phase over 100-300 K (time step: 1 fs; total duration: 10 ns). Through trajectory visualization analysis and fitting of the volume-temperature relationship (Fig.~\ref{fig:phase_transition}), a first-order structural phase transition was identified at approximately 143~K, characterized by discontinuous jumps in the lattice parameters and the abrupt disappearance of octahedral rotation modes. The system potential energy and the in-plane (x and y) pressure components exhibit clear discontinuities at the same temperature, further indicating the occurrence of a structural phase transition. The low-temperature phase ($<$143~K) exhibits conventional positive thermal expansion, whereas the high-temperature cubic phase ($>$143~K) displays significant negative thermal expansion. The calculated volumetric thermal expansion coefficient, $\alpha_v = -14.67 \times 10^{-6}$~K$^{-1}$, is in excellent agreement with the experimental value ($-15.78 \times 10^{-6}$~K$^{-1}$, with a deviation of only ~7\%), validating the accuracy of the NEP in capturing the microscopic mechanism of negative thermal expansion. The simulated transition temperature (143 K) underestimates the experimental value (200 K). This deviation likely arises from limitations of the PBEsol exchange-correlation functional used in generating the training data, as also noted in previous studies\cite{fransson2023limits,fransson2023revealing,fransson2023phase}.

\begin{figure}[ht]
    \centering
    \includegraphics[width=16cm]{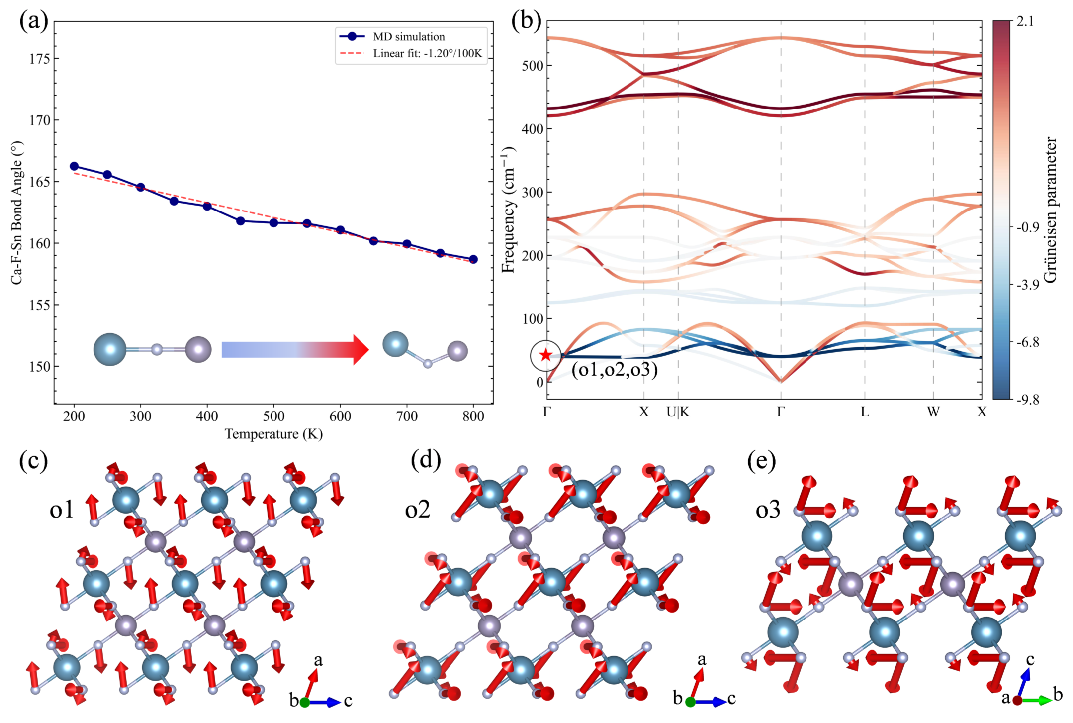}
    \caption{
(a) Temperature dependence of the Ca-F-Sn bond angle.
(b) Phonon dispersion projected with mode-resolved Gr\"uneisen parameters.
(c-e) Vibrational patterns of the three lowest-frequency optical modes ($o_1$, $o_2$, $o_3$) at the $\Gamma$ point.
    }
    \label{fig:phase_transition}
\end{figure}

The negative thermal expansion in CaSnF$_6$ is intimately linked to its structural framework and vibrational dynamics. Similar to canonical NTE materials such as ScF$_3$, ReO$_3$, and ZrW$_2$O$_8$\cite{dove2016negative}, the structure consists of corner-sharing [CaF$_6$]$^{4-}$ octahedra forming a three-dimensional flexible network. This topology supports low-energy rigid unit modes (RUMs), involving cooperative rotations of nearly rigid octahedra. To elucidate the microscopic origin of NTE, Fig.~5(b) shows the phonon dispersion projected with mode-resolved Gr\"uneisen parameters. The three lowest-frequency optical branches ($o_1$, $o_2$, $o_3$) exhibit the largest (negative) Gr\"uneisen parameters, indicating their dominant contribution to NTE. The corresponding vibrational eigenmodes at the $\Gamma$ point are shown in Figs.~5(c-e), where the three modes are nearly degenerate. These modes primarily involve rotational motions of [CaF$_6$]$^{4-}$ octahedra coupled with transverse vibrations of F atoms. Rotation of [SnF$_6$]$^{2-}$ was also observed at other high-symmetry points. Upon heating, the transverse motion of F atoms and the cooperative rotations of [CaF$_6$]$^{4-}$ and [SnF$_6$]$^{2-}$ octahedra (RUMs) induce bending of the Ca-F-Sn bond angle, thereby shortening the Ca-Sn distance while largely preserving the Sn-F bond length, leading to a net volume contraction. This effect is directly illustrated in Fig.~5(a), where the Ca-F-Sn bond angle decreases monotonically with temperature, with a reduction of approximately $1.20^\circ$ per 100 K. These modes possess negative Gr\"uneisen parameters and are thermally activated, giving rise to the observed macroscopic negative thermal expansion.

\begin{figure}[ht]
    \centering
    \includegraphics[width=12cm]{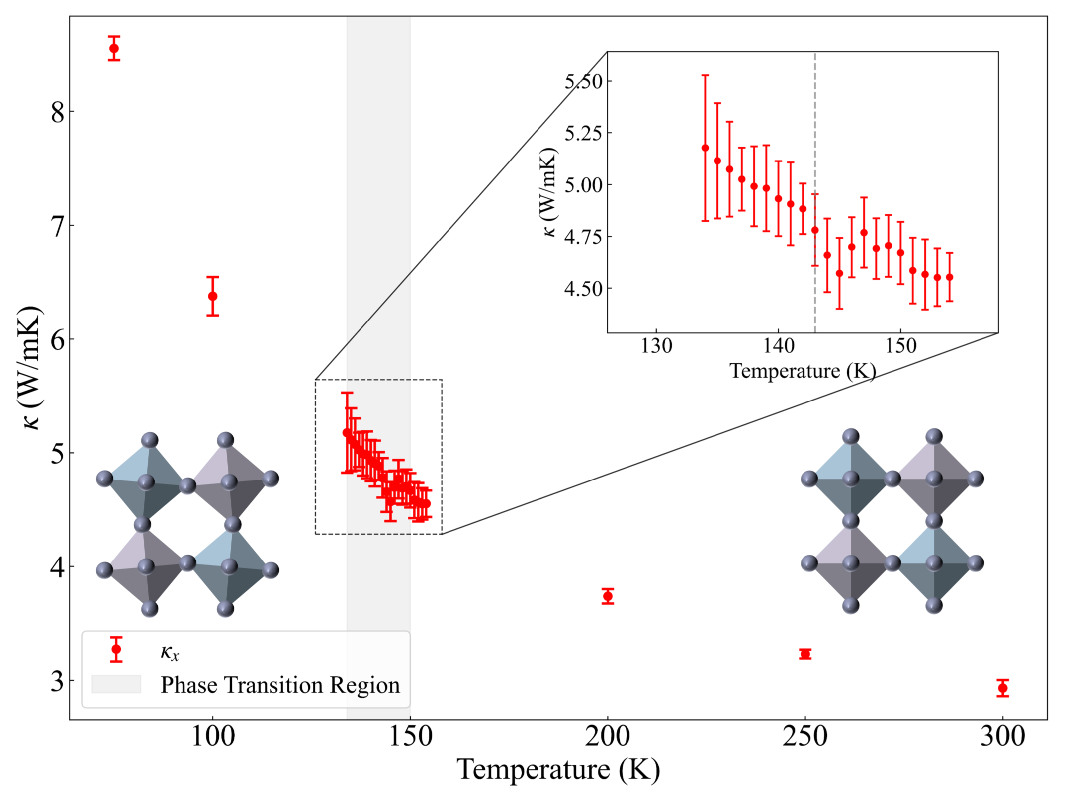}
    \caption{
        Anomalous thermal conductivity behavior of CaSnF$_6$ near the phase transition temperature, which obtained from HNEMD calculations.
    }
    \label{fig:kappa_jump}
\end{figure}

To examine transport properties in the vicinity of the phase transition, we performed HNEMD simulations based on the same NEP potential, tracking the $\kappa_L$ over 75-300 K. As shown in Fig.~\ref{fig:kappa_jump}, the $\kappa_L$ follows an approximate $\sim 1/T^{\alpha}$ dependence away from the transition, consistent with conventional phonon-dominated transport. In contrast, pronounced deviations emerge in the vicinity of the phase transition (142-148 K). The $\kappa_L$ decreases from 4.88 Wm$^{-1}$ K$^{-1}$ at 142 K to a minimum of 4.57 Wm$^{-1}$ K$^{-1}$ at 145 K, followed by a partial recovery to 4.70 Wm$^{-1}$ K$^{-1}$ at 146 K before resuming the normal trend. This non-monotonic behavior-characterized by suppression, minimum, and recovery-coincides with the structural transformation identified in NPT simulations. The initial reduction reflects enhanced phonon scattering associated with lattice instability and structural reconstruction near the transition. The subsequent recovery suggests a rapid reorganization of the phonon spectrum upon completion of the phase transition, leading to a transient improvement in heat transport. As a quantity highly sensitive to lattice structure and vibrational dynamics, the anomalous evolution of $\kappa_L$ provides independent and compelling evidence for the first-order structural phase transition in CaSnF$_6$. Notably, the sharp and structured nature of this anomaly cannot be attributed to statistical or computational uncertainty, but instead reflects the intrinsic reconstruction of phonon transport channels across the phase boundary.

\section{Conclusion}
In summary, we establish a comprehensive picture of the structural phase transition in CaSnF$_6$ and its strong coupling to thermal transport by integrating first-principles calculations with a high-fidelity NEP. The developed NEP accurately reproduces the underlying energy landscape and enables large-scale molecular dynamics simulations that capture the first-order phase transition, bridging the gap between atomistic accuracy and mesoscopic dynamics. From a microscopic perspective, the NTE is shown to originate from low-energy rigid unit modes associated with cooperative rotations of corner-sharing [CaF$_6$]$^{4-}$ and [SnF$_6$]$^{2-}$ octahedra, which induce bond-angle bending and volume contraction. Concurrently, strong anharmonicity, manifested in enhanced 4ph scattering, plays a dominant role in suppressing $\kappa_L$. Crucially, HNEMD simulations reveal a pronounced non-monotonic anomaly in $\kappa_L$ in the vicinity of the phase transition, providing direct transport-level evidence of lattice reconstruction. This anomalous behavior reflects the competition between phonon hardening, scattering enhancement, and rapid reorganization of vibrational states across the phase boundary. Overall, this work uncovers a unified mechanism linking lattice geometry, anharmonic lattice dynamics, and thermal transport in CaSnF$_6$. The combined framework of machine-learned potentials and multiscale transport simulations provides a robust and general strategy for investigating phase-transition-driven functional responses in complex materials.

\begin{acknowledgments}
This research was supported by the National Natural Science Foundation of China under Grants No. 12204402.
\end{acknowledgments}

\bibliographystyle{unsrt}

\end{document}